\documentstyle[12pt,fleqn]{article}
\begin{document}
\indent {SOME COMMENTS ON THE NATURE OF UNIVERSAL PROPERTIES\\
\indent IN LOW--TEMPERATURE  GLASSES\footnote{appears in the
Proceedings of the V International Meeting on Hole-Burning
and Related Spectroscopies; published in 
 Molecular Crystals and Liquid Crystals (1996)}\\

\indent {DAVID R. REICHMAN, PETER NEU,  AND ROBERT J. SILBEY}\\

\indent Department of Chemistry,
 Massachusetts Institute of Technology,\\
\indent  Cambridge, Massachusetts 02139, USA\\

\underline{Abstract} 
We discuss the recent theory of Burin and Kagan that attempts to\\
\indent  explain the existence 
of universal low temperature properties in amorphous \\ \indent solids.
 We suggest a realistic 
experimental scenario that could be used to test \\
\indent the theory.
 We comment on the results 
of an experiment that has already \\
\indent been performed  in the proposed geometry.
\\
\\

\noindent \underline{INTRODUCTION}

\vspace{0.25cm}

\noindent Many different amorphous solids display a remarkable universal behavior at low
temperatures.$^1$  Examples of such behavior include a specific heat and thermal
conductivity that have roughly linear and quadratic temperature dependencies,
respectively, below  about 1K. ${^2}$  In addition to these qualitative similarities, low temperature
amorphous solids show dramatic {\em quantitative} universalities.  An example
of this type of universality is demonstrated in the relation $\frac{l}{\lambda} \sim 150$, where
$l$ is the phonon mean free path and $\lambda$ is the phonon wavelength. For many amorphous
solids, this relation holds to within a factor of 2, below 1K.$^3$

The first type of universal behavior, manifested in the qualitative similarities in the specific heat
and thermal conductivity of a variety of amorphous solids at low temperatures, can be described
by the standard tunneling model.$^4$  In this picture, the glass is viewed as a metastable configuration
of atoms.  In such a configuration, it is possible that an atom or group of atoms may reside in either
of two equilibrium positions.  The potential energy curve for this situation can be represented as a 
double well potential.  At low temperatures, the atom or group tunnels from one equilibrium position
to the other.  In a basis consisting of states localized in the left and right wells, respectively,
the Hamiltonian for the tunneling process may be expressed by
\begin{equation}
H = \frac{1}{2} \left( \begin{array}{clcr} \epsilon & -\Delta_{0} \\
                                           -\Delta_{0} & -\epsilon \end{array} \right).
\end{equation}
Here $\epsilon$ is the asymmetry energy (difference in energy between the left and right wells), and
$\Delta_{0}$ is the tunneling matrix element that connects the lowest energy states in each well.  The
standard tunneling model then dictates that the distributions of the asymmetry energy $\epsilon$ and
 the tunneling matrix element $\Delta_{0}$ are given by 
\begin{equation}
P(\epsilon,\Delta_0) = {\overline{P}\over\Delta_0}
\end{equation}
with a constant $\overline{P}$.
  These assumptions lead directly
to a specific heat that varies linearly with temperature, and a thermal conductivity that varies as
$T^{2}$.$^4$

Various aspects of the standard tunneling model may be questioned.  First, the microscopic foundation
for the model is not firmly justified.  While there has been some recent success in ``locating" the
tunneling systems in computer models of disordered solids,$^5$ questions still remain.  Though the uniform
distribution in the asymmetry energy is quite reasonable, there is no firm justification for the flat
distribution in $\log (\Delta_{0})$.  Furthermore, the standard tunneling model cannot explain the remarkable
{\em quantitative} universality in the ratios of certain parameters, for instance the relation $\frac{l}{\lambda} \sim 150$
as explained above.$^6$  As a result of these inadequacies, several alternative models have been proposed.$^{6,7}$
A common theme in these models is the belief that the interactions between the tunneling systems dominate
the energy scale at low temperatures.  Recently, Burin and Kagan$^8$ have devised a model that attempts to explain
not only the form of the distribution of tunneling center parameters, but also the quantitative resemblance
of certain properties observed in various glasses.  In this note, we will briefly discuss the salient features
of the model of Burin and Kagan.  We will then propose a realistic experimental scenario that we believe can
be used to test their model.  Lastly, we briefly comment on one experiment that has already
been performed that might shed some light on this issue.

\vspace{1cm}

\noindent \underline{THE MODEL OF BURIN AND KAGAN}

\vspace{0.25cm}

\noindent In the model of Burin and Kagan$^8$, the strain mediated dipole-dipole interaction between tunneling centers
is responsible for the universal properties observed in low temperature glasses.  The amorphous medium consists
of double well centers distributed randomly in space with an arbitrary distribution of parameters. Unlike 
the situation described in standard tunneling model, the parameters describing the randomly distributed 
tunneling centers, called ``primary defect parameters'', do not display universal behavior.  That is, the 
distribution of the defect energy is not necessarily flat, and will vary depending on the chemical composition
of the glass. The universal properties appear as a consequence of the many body interaction of the primary 
tunneling systems, leading to the creation of delocalized excitations called ``many center excitations".  
Due to the delocalization, the 
spectral properties of the many center excitations are independent of the primary defect parameters 
(thus leading to universal ratios such as $\frac{l}{\lambda} \sim 150$)  and demonstrate practically 
uniform distributions in the energy asymmetry and the logarithm of the tunneling matrix element.  The role
of the many center excitations increases with decreasing energy (temperature).  

The crucial aspect in the formation of the many center excitations is the fact that the strain mediated 
interaction between the primary tunneling centers decays as $1/R^{3}$.  In three dimensions,  
 the average number of primary tunneling centers
forming a multicenter excitation increases logarithmically with glass volume
\begin{equation}\label{log}
N(V)\sim \alpha \,\log(V)\ .
\end{equation}
  This logarithmic
behavior allows Burin and Kagan to study the formation of many center excitations with a renormalization
group approach.  A similar procedure was first used by Levitov$^9$ in the study of the  delocalization
of vibrational modes caused by the electric dipole interaction.
The logarithmic divergence of $N(V)$ indicates {\it criticality} ($N(V)\sim V^\alpha$),
 and, hence, {\it delocalization}.$^9$ According to this argument, a ``modified dipole-dipole interaction'',
in three dimensions, $1/R^{3+\eta}$,    prohibits  the formation of delocalized multicenter excitations
for $\eta > 0$. As a result, the distribution of primary tunneling centers gains importance  and,
following the argument of Burin and Kagan, no universal behavior of glasses is expected. In contrast,
for $\eta < 0$ the number of primary centers forming a delocalized excitation diverges.

While we will not recapitulate the detailed arguments of Burin and Kagan, we would like to highlight 
some important features of their argument.  Starting from the Hamiltonian
\begin{equation}\label{H}
H = -\sum_{i} \omega_{i}S_{i}^{z} -\frac{1}{2} \sum_{ij}U_{ij}S_{i}^{z}S_{j}^{z} - \sum_{i}\Delta_{0i}S_{i}^{x},
\end{equation}
where $\omega_{i}$ is the asymmetry energy  (previously referred to as $\epsilon$),
$\Delta_{0i}$ is the  tunneling matrix element of the $i$th primary tunneling center, and $U_{ij} = \frac{u_{ij}}{R^{3}_{ij}}$
gives the interaction strength between primary centers.  The asymmetry energy is distributed in the interval
$(-W/2, W/2)$, the average scale for the tunneling amplitudes of the primary centers is $\Delta_{0 *}$, and
$U_{0} = \langle |u_{ij}| \rangle$ gives the characteristic scale for interactions of the primary centers.
In the standard tunneling model, the parameters $\omega_i$ and $\Delta_{0i}$ have preassigned distributions, whereas
here Burin and Kagan assume no specific form for the distribution of these primary defect parameters.  It is,
however, assumed that 
\begin{equation}
W \gg U_{0}n \gg \Delta_{0 *}
\end{equation}
where $n$ is the density of tunneling centers.  This
allows Burin and Kagan to first neglect the tunneling, and to include its effects in a perturbative fashion
in the parameter $\Delta_{0*}/W$. 

Now the density of states for the asymmetry energy in the presence of TLS-TLS interactions, $P(\Delta)$, is
considered (here $\Delta$ is the asymmetry energy modified by the interactions).  At zero 
temperature, the system should be in the ground state.  This means that the energies of
the multicenter excitations should be positive.  For single particle excitations, neglecting for now the 
last term in the Hamiltonian (\ref{H}), this fact is embodied in
the stability criteria 
\begin{equation}
 \Delta_{i} =\omega_{i} + {1\over 2}\sum_{j}U_{ij}S^{z}_{j} > 0 \ .
\end{equation}
 This type of stability condition
may be extended to include $n$ centers.  For example, pair excitations have the stability requirement
\begin{equation}
\Delta_{ij} = \Delta_{i} + \Delta_{j} -U_{ij} > 0\ ,
\end{equation}
 three center interactions $\Delta_{ijk} =
\Delta_{i} + \Delta_{j} + \Delta_{k} -U_{ij} -U_{ik} -U_{jk} > 0 $ and so on.  Burin and Kagan first 
consider a restricted range of interaction, $R_{0}$, limited enough to consider the intercenter interactions 
as a weak perturbation.  Next, the decrease in the density of single particle excitations caused by the 
stability criteria for pair excitations is calculated,
\begin{equation}
P_{1}(\Delta) = \frac{1}{V} \sum_{i} \langle \delta(\Delta - \Delta_{i}) \prod_{i} \theta(\Delta_{ij}) \rangle.
\end{equation}
Here, $V$ is the system volume, and $\theta(\Delta_{ij})$ is a step function that enforces the stability
requirement for pair excitations.  The density of single particle excitations is then approximated as 
\begin{equation}
P_{1}(\Delta) \sim P_{0} \left(1-P_{0}\int d {\bf R_{12}} \int d \Delta_{1} \langle \theta ( \frac{u_{ij}}{R^{3}_{12}} -
\Delta_{1} - \Delta) \rangle_{u} \theta(R_{0} - R_{12}) \right),
\end{equation}
where $\langle ... \rangle_{u}$ denotes an average over the $u_{ij}$.  Burin and Kagan assume that $P(\Delta)$
has no singularity at $\Delta =0$, allowing for the replacement of $P(\Delta_{1})$ with $P_{0} = P(0) \approx
n/W$ because the main contribution to the above integral comes from small values of $\Delta_{1}$.  The above 
integral may be performed, yielding
\begin{equation}
P_{1}(\Delta) \approx P_{0}(1-2\chi \xi),
\end{equation}
where 
\begin{eqnarray}
\chi &=& \pi P_{0} U_{0} \ll 1 \\
\xi &=& \ln(R_{0}/R_{min})
\end{eqnarray}
 with $R_{min}$ defined through
$U_{0}/R^{3}_{min} \approx W$. A similar calculation for the density of states for pair excitations yields
\begin{equation}
P_{2}(\Delta) \approx P_{0}\chi \xi
\end{equation}
and, in general, for many center excitations, $P_{n} \sim (\chi \xi)^{n-1}$.  The crucial point is that as $R_{0}$
increases, the product $\chi \xi$ becomes larger ($\chi \xi \sim\mbox{$\cal{O}$}(1)$), signaling the decrease in the 
importance of the single particle (primary tunneling center) properties, and the onset of many center excitations.
Using a renormalization group approach, Burin and Kagan calculate  the density of states $P(\Delta,\Delta_{0})$
including the influence of the many body terms.  They find, after tunneling effects are included, a distribution that may be 
approximately written as $P(\Delta,\Delta_{0}) = \overline{P}/\Delta_{0}$ with a value of 
$ C \equiv \overline{P} U_{0} \sim 10^{-3}-10^{-4}$ that is 
in agreement with experiments.$^3$  This distribution results from the consideration of many body effects;
 {\em the primary (noninteracting) set of tunneling systems do not show this universal behavior}.  The universal parameter
 $C$ appears due to the delocalization caused by the 
TLS-TLS interaction that produce many center excitations, effectively ``washing out" the details of the chemical
structure of the particular glass under study.

\vspace{1cm}

\noindent\underline{A PROPOSED TEST}

\vspace{0.25cm}

\noindent A crucial aspect of the theory of Burin and Kagan is the long range $1/R^3$ dipole-dipole force between
the TLS resulting in delocalization. This fact 
is responsible for universal low temperature properties in glasses. 
This was originally conjectured by Yu and Leggett.$^6$
Following the reasons given below Eq. (\ref{log}), one has to put the 
system out of criticality in order to test this conjecture.
One may do this by either changing the form of the TLS interaction 
or by confining the spatial geometry of the primary tunneling centers.

First work in the latter  direction has been done by Fu.$^{10}$
He proposed a study
of the properties of a free-standing, thin glass wire.  He showed that in a thin fiber of radius $R_{*}$, the
long ranged $1/R^{3}$ force is modified to $U \sim \exp(-R/R_{*})$ for two defects separated by a distance
$R > R_{*}$.  Thus, the hypothesis that long range forces are responsible for the universal properties observed
in glasses may be tested in a thin glass fiber, where the dipole-dipole force is no longer long ranged.

As far as we know, no experimental studies have been made on such a system.  The reasons for this are
twofold.  First, it is very difficult to do low temperature studies on free fibers.  
 Here, the coat surrounding the fiber will greatly increase the fiber radius, and the need
for good thermal contact with a refrigerator will make isolation of the fiber difficult (or impossible).  Second,
the temperature must be extremely low in order to see the effects of the confining wire.  One recent estimate of the
temperature needed to see the effects proposed by Fu is $T \sim 10^{-7}$~K
 for a wire with a $1 \mu m$ diameter.$^{11}$

Is it possible to test, in a realistic way, theories like the one outlined in the section above?  We believe
the answer is yes. 
  The crucial point to note in the theory of Burin and Kagan is the need for
a TLS-TLS coupling that varies as $1/R^{3}$ {\em in three dimensions}.  The story is drastically
different if all the TLS are confined to quasi-two dimensions, while the coupling between them still varies 
as $1/R^{3}$.  Consider a layer of thickness $a$, where $a$ is of atomic dimensions.  If such
a layer is glassy, and lies on a substrate that is thick and contains no TLS dynamics,  then we are approximately
in the regime where the TLS dynamics are confined to two dimensions, while their interaction still varies
as $1/R^{3}$.  
We implicitly assume that sound waves are not affected by the interface between
the amorphous layer and the bulk.
To see how this situation varies from the usual one, consider the parameter $\xi$ of the 
last section for $R\gg a$
\begin{equation}
\xi = {1\over 4\pi} \int \frac{d {\bf R_{ij}}}{R^{3}_{ij}} 
\approx {a \over 2}  \left(\frac{1}{R_{min}} - \frac{1}{R_{0}} \right).
\end{equation}
Since $R_{min}$ is of the order of the size of the primary tunneling centers,
the parameter $\xi$ is always  ${\cal O}(1)$, and never shows the logarithmic  growth
characterized by the usual situation in three dimensions.  
Accordingly, in the geometry proposed above, the parameter $\chi \xi$
defined in the last section will {\em always be small}, obviating the importance of the many center excitations.
In such a case, the intrinsic ``primary" distributions should dominate, and universal properties will
be lost (if one accepts the arguments of Burin and Kagan).

 The experiment we  suggest   has already been performed, albeit not for the
purpose that we discuss here.
The  hole burning experiment of Orrit, Benard, and M\"{o}bius on an ionic dye in a Langmuir-Blodgett
monolayer is an experiment of the type we propose above.$^{12}$  In this experiment, persistent holes in the
excitation spectrum of resorufin adsorbed on an ammonium salt monolayer were measured.  The monolayer
is disordered due to preparation effects and the holes showed signatures of glassy behavior.  In fact,
Pack and Fayer$^{13}$ were able to 
explain this data qualitatively  by assuming a standard tunneling model
description of monolayer.  {\em This contradicts  the theory of Burin and Kagan}, which (as we
have pointed out above)  predicts results at variance with the standard model in a two dimensional
system with $1/R^{3}$ TLS-TLS interactions.  It would be important, however, for a variety of such experiments
to be performed with different ``glassy'' monolayers, so that a definitive conclusion can be reached.

\vspace{1cm}

\noindent {\underline{ACKNOWLEDGEMENTS}

\vspace{0.25cm}

\noindent We would like to thank the National Science Foundation for partial support of this research.  
One of us (P.N.) would like to thank the Alexander von Humboldt Foundation for financial support.  
We would also like to thank Professor L. Levitov for useful discussions.

\vspace{1cm}

\noindent\underline{REFERENCES}
\begin{enumerate}
\item  For a review, see: \underline{Amorphous Solids -- Low Temperature
             Properties},\\
            \underline{Topics in Current Physics},  \underline{24},
             edited by W. A. Phillips (Springer, Berlin Heidelberg New York,
             1984).
\item   R.C. Zeller and R.O. Pohl, \underline{Phys. Rev. B}, \underline{4}, 2029 (1971).
\item  J.J. Freeman and A.C. Anderson, \underline{Phys. Rev. B}, \underline{34} 5684 (1986).
\item  P.W. Anderson, B.I. Halperin and C.M. Varma, \underline{Philos. Mag.}, \underline{25} 1 (1972); 
       W.A. Phillips, \underline{J. Low Temp. Phys.}, \underline{7} 351 (1972).
\item  A. Heuer, R.J. Silbey, \underline{Phys. Rev. Lett.}, \underline{70} 3911 (1993).
\item  C.C. Yu and A.J. Leggett, \underline{Comments Condensed Matter Phys.}, \underline{14}, 231 (1991).
\item  S.N. Coppersmith, \underline{Phys. Rev. Lett.}, \underline{67} 2315 (1991).
\item  A.L.Burin and Yu. Kagan, \underline{Phys. Lett. A}, \underline{215} 191 (1996); 
       A.L. Burin and Yu. Kagan, \underline{JETP}, \underline{82} 159 (1996).
\item   L.S. Levitov, \underline{Europhys. Lett.}, \underline{9} 83 (1989); L.S. Levitov, 
        \underline{Phys. Rev. Lett}, \underline{64} 547 (1990).
\item   Y. Fu, \underline{Phys. Rev. B}, \underline{40} 10056 (1989).
\item   S.N. Coppersmith, \underline{Phys. Rev. B}, \underline{48} 142 (1993).
\item   M. Orrit, J. Bernard, and D. M\"{o}bius, \underline{Chem. Phys. Lett.}, \underline{156} 233 (1989).
\item   D.W. Pack and M.D. Fayer, \underline{Chem. Phys. Lett.}, \underline{168} 371 (1990).

\end{enumerate}
\end{document}